\documentclass[11pt]{article}
\usepackage{amsmath}
\usepackage[backend=biber,style=apa,sorting=nyt]{biblatex}
\usepackage[colorlinks=true, allcolors=blue]{hyperref}
\usepackage{cleveref}
\usepackage{csquotes}
\usepackage{lineno}
\usepackage{amsopn}
\usepackage{orcidlink}
\usepackage{authblk}
\usepackage[utf8]{inputenc}
\usepackage[T1]{fontenc}
\usepackage{geometry}
\usepackage{graphicx}
\usepackage{caption}
\usepackage{subcaption}
\usepackage{textcomp}
\usepackage{amsfonts}
\usepackage{amssymb}
\usepackage{booktabs}
\usepackage{setspace}
\usepackage{tikz}
\usetikzlibrary{calc, angles, quotes, intersections, arrows.meta, decorations.markings, decorations.pathreplacing}
\MakeOuterQuote{"}
\DeclareNameAlias{sortname}{family-given}
\DeclareNameAlias{default}{family-given}
\DeclareCiteCommand{\posscite}
  {\usebibmacro{prenote}}
  {\ifciteindex
     {\indexnames{labelname}}
     {}%
     \printnames{labelname}%
     \printtext{'s}%
     \addspace
     \printtext[parens]
        {\printtext[bibhyperref]{\printfield{year}}}}
  {\multicitedelim}
  {\usebibmacro{postnote}}

\title{Quantifying Avian Morphological Evolution through Deep Representation Learning} 

\author[1,2]{Jiao Sun\orcidlink{0000-0002-5028-8132}}
\affil[1]{Division of Ecology and Evolutionary Biology, School of Biological Science, University of Reading, Whiteknights, Reading, RG6 6EX, United Kingdom}
\affil[2]{CAS Key Laboratory of Plant Germplasm Enhancement and Specialty Agriculture, Wuhan Botanical Garden, Chinese Academy of Sciences, Wuhan 430074, China}
\affil[ ]{\url{j.sun@pgr.reading.ac.uk}}
\affil[ ]{\url{sunjiao19@mails.ucas.ac.cn}}

\date{\today}

\begin{document}

\maketitle

% \doublespacing

% \linenumbers

\begin{abstract}
The evolution of biological morphology is fundamentally linked to ecological adaptation and species survival, yet traditional morphological evolution relies on landmark-based geometric morphometrics, a process constrained by subjective manual annotation, strict requirements for anatomical homology, and an inability to easily quantify complex, non-rigid traits such as plumage and texture. To overcome these limitations, we propose a scalable, landmark-free morphometric framework driven by deep learning. By extracting high-dimensional feature vectors from a Convolutional Neural Network (ResNet34) trained on images of over 10,000 bird species, we project raw visual semantics into a high-dimensional morphospace. Even without a priori taxonomic knowledge, this visual morphospace naturally recovers classical hierarchical taxonomy and effectively captures both homology and convergence. Analyses reveal a highly significant phylogenetic signal within the network's embeddings, with principal components correlating strongly with established ecological and morphological traits. Furthermore, by implementing a novel spherical Ancestral State Reconstruction algorithm, we uncover a pronounced "early-burst" pattern of disparity following the K-Pg mass extinction, supporting the niche-filling hypothesis of adaptive radiation.
\end{abstract}

\addcontentsline{toc}{section}{Abstract}


\section{Introduction}\label{sec:introduction}

The evolution of biological morphology plays a crucial role in shaping
the diverse natural world we observe today. It provides insight into
the adaptation and survival of species over time, influencing various
ecological interactions and the functioning of ecosystems.
Traditionally, analyses of morphological evolution require human intervention in
the selection and coding of traits. 
The primary limitation of landmark-based geometric morphometrics is its
dependence on subjective human selection and the strict anatomical comparability of the chosen features. 
Before any mathematical analysis can occur, researchers must manually define, physically or digitally locate, 
and mathematically encode numerous landmarks across every single specimen within a given dataset. 
This manual pipeline is highly susceptible to human bias.
Furthermore, geometric morphometrics is almost exclusively optimized for the analysis of hard, rigid, 
static anatomical shapes -- primarily skeletal osteology. Many soft features, like complex textures exhibited
in many taxa, are hard to measure and encode in traditional morphometrics. These traits are 
also critical to ecological adaptation or sexual selection. 
The inability of traditional methods to quantify these highly complex, non-rigid phenotypic traits 
results in an underestimate of morphological disparity \parencite{daboul2018procrustes, Guillerme2020Disparities, mulqueeney2025assessing}.

To address these limitations, we propose an advanced deep
learning technology. Convolutional Neural Networks, 
popularised by \textcite{LeCun1989Backpropagation},
are designed to automatically learn visual features from image data,
making them exceptionally suited for visual recognition tasks. By utilising
CNN image classification models, we can leverage the learned weights
as indicators of morphological traits of various species, providing a
more objective basis for understanding biological evolution.
It can serve as a high-throughput, high-dimensional
morphometric tool capable of capturing holistic visual features, including colour,
texture, and body plan. 

To validate the feasibility of this proposed methodology, 
we trained a model being able to recognise over 10,000
species of birds, based on ResNet34 architecture \parencite{He2016ResNet},
and calculated the cosine similarity between species using the weights
extracted from the last fully connected layer (fc). This methodology
enabled us to perform phenotypic disparity and macroevolutionary analysis,
yielding insights into the morphological evolution of birds.

% To avoid being confused with the taxonomic level "class," the term "category"
% is used as an alternative to the "classes" of neural networks.

\section{Material \& Methods}\label{sec:masterial-methods}

\subsection{Data processing}\label{subsec:mm-data-processing}

In this case study, we utilised the DongNiao International Birds 10000 Dataset (DIB-10K)
by \textcite{Mei2020Dongniao}, which comprises over 4.8 million images representing
10,922 bird species, following the IOC 10.1 taxonomy \parencite{Gill2021IOC}. This
extensive dataset encompasses a wide variety of bird species, morphological variants,
postures, and gestures \parencite{Mei2020Dongniao}. Despite their efforts in manual
review and correction, the dataset contains numerous duplicate and erroneous images,
necessitating a comprehensive data cleaning process.

For the data cleaning, we employed the pHash method to detect near-duplicate images
by comparing these hashes \parencite{Farid2021Overview}. Duplicates spanning multiple
categories are completely removed, and for intracategory duplicates, only one copy is
retained in its category.

To address erroneous images with non-avian subjects, we utilised the pretrained Faster
Region-based Convolutional Neural Network (Faster R-CNN,~\cite{Ren2017Faster}) in
the torchvision library \parencite{Ansel2024Pytorch}. For images where no birds were
detected, we implemented a two-tiered approach: images from high-volume categories
($\geq200$ images) were automatically deleted, while those from low-volume categories
($<200$ images) were moved to a separate directory for manual inspection.
After a manual thorough review, valid images were reintegrated into the dataset.

\subsection{Model training}\label{subsec:mm-model-training}

In this step, we project the raw visual semantics of organisms into a mathematically tractable morphological space. 
We define a mapping function $f_\theta$ parameterised by a Convolutional Neural Network (CNN). For a given image of an organism, the network directly processes the pixel matrix and extracts a high-dimensional feature vector from its final fully connected (fc) layer, bypassing the requirement for homologous anatomical coordinates.

The task of recognising over 10,000 bird species is a
fine-grained visual categorisation (FGVC) problem, where the goal is
to identify minor differences between highly similar categories. To
tackle this challenge, we adopted the MetaFGNet model proposed by \textcite{Zhang2018Fine},
which is specifically designed for FGVC tasks. We utilised their
pretrained weights "LBird-31\_checkpoint" as the foundation for
transfer learning on the processed DIB-10K dataset. In the training process,
all species are regarded as equal categories, without any \textit{a priori} taxonimic
knowledge being introduced.

For the training process, we employed a server equipped with an RTX 4090D
GPU, 90GB of memory, and a 15-core
Intel\textsuperscript{\textregistered} Xeon\textsuperscript{\textregistered}
Platinum 8474C processor from the \href{https://www.autodl.com/docs/}{AutoDL platform}.
This high-performance setup allowed for efficient processing and model training.
The training procedure was conducted for 32 epochs.

We reassigned the orders and families of all species (categories) to align with the
IOC 15.1 \parencite{Gill2025IOC}. The weights are extracted from the final fully
connected layer (fc) of the ResNet34 model. Weights of each species were regarded as a
512-dimensional vector representing the morphological traits. They are reduced to the
lowest boundary that can explain 80\% of all variances for subsequent analyses.
Gradient-weighted Class Activation Mapping (Grad-CAM,~\cite{Selvaraju2020Grad}) is
employed to visualise the model's attention on images.

% The extracted vector is subsequently $L_2$-normalised. Every organism is mathematically represented as a unit vector $v$. 
% This normalisation forces the deep visual morphospace to function geometrically as a unit hypersphere $\mathbb{S}^{D-1}$, 
% where $D$ is the dimensionality of the feature space. Within this Riemannian manifold, semantic and morphological similarity is encoded in angular directions rather than spatial magnitudes.

\subsection{Geometric Foundations of the Deep Morphospace}\label{subsec:mm-geometric-foundations}

To employ classical macroevolutionary models (such as Brownian motion and Phylogenetic Independent Contrasts) 
on feature representations extracted from deep neural networks, it is crucial to define the precise geometry of the underlying feature space. 
As demonstrated by \textcite{Wang2017NormFace} in the context of hyperspherical embedding, the loss function with $L_2$ normalization forces 
the network to minimize intra-class angular variance while maximizing inter-class angular disparity. 
Consequently, the semantic and morphological information learned by the model is primarily encoded in the angular direction of the feature vectors, 
rather than their Euclidean magnitudes.

Mathematically, the $L_2$ normalisation maps any vector $\mathbf{x} \in \mathbb{R}^D$ onto a unit hypersphere $\mathbb{S}^{D-1}$, defined as:

$$\mathbf{z} = \frac{\mathbf{x}}{\|\mathbf{x}\|_2}$$

where $\|\mathbf{z}\|_2 = 1$. 
Therefore, the deep visual morphospace in our framework does not function as a flat Euclidean space, 
but rather as a curved Riemannian manifold -- specifically, a unit hypersphere. 
The phenotypic distance between any two taxa $i$ and $j$ cannot be measured by Euclidean distance, which would violate the constraints of the surface topology. 
Instead, it is quantified via the geodesic distance (the great-circle distance) or the angular separation $\theta_{ij}$, formulated as:

$$\theta_{ij} = \arccos(\mathbf{z}_i \cdot \mathbf{z}_j)$$

This hyperspherical formulation provides the theoretical and geometric justification for our subsequent developments in spherical 
Ancestral State Reconstruction (ASR) and Riemannian Phylogenetic Independent Contrasts (PIC). 

\subsection{Similarity clustering}\label{subsec:mm-similarity-clustering}

According to the aforementioned geometric foundations, species similarity can be quantified using cosine similarity, implemented via
dot product calculation of $L_2$-normalised vectors in PyTorch \parencite{Ansel2024Pytorch}.
Next, we conducted agglomerative hierarchical clustering using the average
linkage method to merge clusters. This was executed with the
hierarchical clustering functionalities implemented in the SciPy
library \parencite{Gommers2025Scipy}. The hierarchical structure of the
clusters was output in Newick format, a widely used format in
computational biology for tree structures. Finally, we utilised ETE3
to export the clustering dendrogram in SVG format \parencite{Huerta2016ETE}.

To analyse the clustering result, we applied a recursive top-down
analysis to the tree, evaluating each internal node for taxonomic
"purity." For a given node, we defined taxonomic purity as
the proportion of the majority taxon. A node with a taxonomic
purity of more than 85\% was considered taxonomically consistent. For
taxonomically consistent nodes, we further examine
all species to identify and annotate taxonomical outliers, which belong
to taxa that different from the majority taxon of this branch.
Finally, we conducted manual review to confirm whether outliers had
biological similarity with the majority taxa of their branches and
what kinds of similarity do they have. 
% The above pipeline was carried out in both order-level and family-level.

\subsection{Disparity analysis}\label{subsec:mm-disparity-analysis}

The Spearman's rank correlation was employed to examine
the relationship between diversity (species richness) and morphological
disparity (spherical variance), where taxa with $N<2$ were excluded. The Akaike
information criterion (AIC) was employed to assess and compare four models:

\begin{enumerate}
    \item Power law: $f(x) = 1 - x^b$, the coefficient of $x^b$
        is fixed as 1 to satisfy the boundary condition $f(1) = 0$,
        reflecting the theoretical expectation;
    \item Stretched exponential model: $f(x) = 1 - \exp(-\lambda \cdot (x - 1)^\beta)$;
    \item Hill equation: $f(x) = \frac{(x - 1)^n}{k + (x - 1)^n}$;
    \item Logarithmic rational model: $f(x) = \frac{\ln(x)}{k + \ln(x)}$.
\end{enumerate}

All analyses were conducted at both order and family levels. Subsequently, the residuals
of the disparity of every taxon are calculated, five taxa with the highest and five with
the lowest residuals are listed.

Additionally, we explore the relationship between taxa size and mean pairwise angle,
taxa size and pairwise angle variance, as well as mean pairwise angle and pairwise
angle variance at both order and family levels. For size vs variance analysis,
taxa with $N<3$ were excluded to avoid mathematical artefacts
($\text{variance} = 0$ for $N=1 \text{or} 2$). For the other two analysis (mean angle),
taxa with $N<2$ were excluded.

\subsection{Phylogenetic tree}\label{subsec:mm-phylogenetic-tree}

The phylogenetic tree used in this study was sourced from \textcite{Stiller2024Complexity}, which encompasses 198 species. The tree is ultrametric, with branch lengths representing divergence times in millions of years (Ma). We extracted a subset containing only the species present in \posscite{Stiller2024Complexity} tree for analyses involving phylogenetic comparative methods, ensuring a consistent set of taxa. 

\subsection{Phylogenetic signal}\label{subsec:mm-phylogenetic-signal}

We employed the R package geomorph to calculate the phylogenetic signal of the embedded vectors. It is represented by the multivariate Blomberg's $ K $ ($ K_{mult} $), being calculated using the geomorph::physignal function. The $ K_{mult} $ = 1 indicating similar levels of phylogenetic signal as expected under a Brownian motion model, $ K_{mult} $ < 1 indicating weaker phylogenetic signal, and $ K_{mult} $ > 1 indicating stronger phylogenetic signal than Brownian motion \parencite{adams2014generalized}. 

The function geomorph::physignal applies Phylogenetically Aligned Component Analysis (PACA) on the provided morphological data. As noted by \textcite{collyer2021phylogenetically}, the number of variables have no significant effect on the estimation of phylogenetic signal. Therefore, the PCA results explaining 100\% variance (198 dimensions) were used for the calculation of phylogenetic signal. The significance of the phylogenetic signal was assessed using 999 permutations, where the species labels were randomly shuffled across the tips of the phylogenetic tree to generate a null distribution of $ K_{mult} $ values.

Furthermore, we calculated the correlation between the Phylogenetically Aligned Components (PACs) and AVONET traits \parencite{tobias2022avonet}. Correlation analyses were conducted using the stats.pearsonr and stats.kruskal functions from the SciPy package \parencite{Gommers2025Scipy}. Pearson correlation was applied to evaluate the relationship between continuous traits and PACs, while the Kruskal-Wallis test was used for categorical traits, with the effect size subsequently quantified as Eta-squared ($\eta^2_H$). To ensure comparability across different trait types, the Pearson coefficient of determination ($R^2$) was used instead of the correlation coefficient ($r$), aligning both metrics to represent the proportion of explained variance. The correlation results were visualised as heatmaps using the seaborn package \parencite{Waskom2021Seaborn}.

\subsection{Disparity through time}\label{subsec:mm-disparity-through-time}

In classical phylogenetic independent contrast (PIC) proposed by
\textcite{Felsenstein1985PIC}, The evolution of phenotypic traits
follows a Brownian motion model in Euclidean space, in which the trait of each descendant
is derived from the trait of the ancestral species through a random walk in Euclidean space.
The time of this walk is proportional to the divergence time, which is the branch length of
the evolutionary tree.

% As demonstrated by \textcite{Wang2017NormFace}, in the last fully connected layer of CNNs,
% the semantic information is primarily encoded in the angular direction, which is equivalent
% to $L_2$ normalised vectors. Therefore, we assume that the feature vectors randomly walk on
% a unit hypersphere.
We implemented the algorithm for Spherical Ancestral State Reconstruction by modeling
phenotypic evolution as Riemannian Brownian Motion on the hypersphere of
the deep feature space in Python.

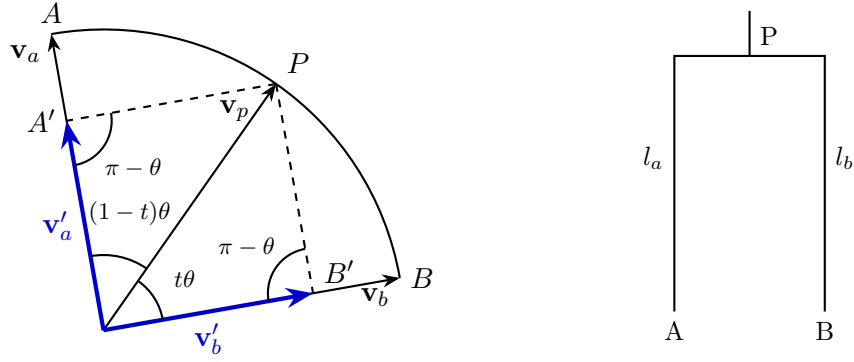
\begin{figure}[htbp]
    \centering
    \begin{minipage}{0.45\textwidth}
        \centering
        \begin{tikzpicture}[>=Stealth, thick, scale=0.8]
            \useasboundingbox (-1.5, -0.5) rectangle (1.5, 5.5);

            % 1. Define Coordinates
            \coordinate (O) at (0,0);
            \coordinate (A_dir) at (100:6); % Direction of Va
            \coordinate (B_dir) at (10:6);  % Direction of Vb

            % Define the main radius
            \def\R{5}
            \coordinate (A) at (100:\R);
            \coordinate (B) at (10:\R);

            % Define P at an interpolated angle
            % Let's assume t is roughly 0.4 for visual similarity
            \coordinate (P) at (55:\R);

            % 2. Calculate Intersections for Parallelogram
            % We need A' on OA such that PA' // OB
            % We need B' on OB such that PB' // OA

            % Path for OA and OB
            \path[name path=rayOA] (O) -- (A_dir);
            \path[name path=rayOB] (O) -- (B_dir);

            % Path from P parallel to OB (to find A')
            \path[name path=lineFromP_parOB] (P) -- ++($(O)-(B_dir)$) -- ++($(B_dir)-(O)$);

            % Path from P parallel to OA (to find B')
            \path[name path=lineFromP_parOA] (P) -- ++($(O)-(A_dir)$) -- ++($(A_dir)-(O)$);

            % Find intersections
            \path [name intersections={of=rayOA and lineFromP_parOB, by=A_prime}];
            \path [name intersections={of=rayOB and lineFromP_parOA, by=B_prime}];

            % 3. Draw Geometry

            % Draw the main arc
            \draw (A) arc (100:10:\R);

            % Draw Vectors OA, OB, OP
            \draw[->] (O) -- (A) node[below left] {$\mathbf{v}_a$};
            \draw[->] (O) -- (B) node[below left] {$\mathbf{v}_b$};
            \draw[->] (O) -- (P) node[below left, xshift=-6pt, yshift=-2pt] {$\mathbf{v}_p$};

            % Draw Components (A' and B')
            % Vectors Va' and Vb' lie on the axes
            \draw[->, ultra thick, blue!80!black] (O) -- (A_prime) node[midway, left] {$\mathbf{v}_a'$};
            \draw[->, ultra thick, blue!80!black] (O) -- (B_prime) node[midway, below] {$\mathbf{v}_b'$};

            % Draw Dashed Parallel Lines
            \draw[dashed] (P) -- (A_prime);
            \draw[dashed] (P) -- (B_prime);

            % 4. Labels and Angles

            % Points labels
            \node[left] at (A_prime) {$A'$};
            \node[above right] at (B_prime) {$B'$};
            \node[above] at (A) {$A$};
            \node[right] at (B) {$B$};
            \node[above right] at (P) {$P$};

            % Angle arcs

            % Angle (1-t)theta between OA and OP
            \pic [draw, angle radius=1.0cm, pic text={$(1-t)\theta$}, angle eccentricity=1.6, font=\footnotesize] {angle = P--O--A};

            % Angle t*theta between OP and OB
            \pic [draw, angle radius=0.8cm, pic text={$t\theta$}, angle eccentricity=1.6, font=\footnotesize] {angle = B--O--P};

            % Internal angles (pi - theta) at A' and B'
            % Using small arcs inside the parallelogram
            \pic [draw, angle radius=0.6cm, pic text={$\pi-\theta$}, angle eccentricity=1.8, font=\footnotesize] {angle = O--A_prime--P};
            \pic [draw, angle radius=0.6cm, pic text={$\pi-\theta$}, angle eccentricity=1.8, font=\footnotesize] {angle = P--B_prime--O};

            % Angle t*theta at P (alternate interior) - optional based on diagram complexity
            % \pic [draw, angle radius=1.0cm, "$t\theta$"] {angle = A_prime--P--O};

        \end{tikzpicture}
    \end{minipage}
    \hfill
    \begin{minipage}{0.45\textwidth}
        \centering
        \begin{tikzpicture}[scale=2,every node/.style={font=\small},line width=0.8pt]
            \coordinate (Root) at (0,0);
            \coordinate (Internal) at (0, -0.3);
            \coordinate (A) at (-0.5, -2);
            \coordinate (B) at (0.5, -2);

            \draw (Root) -- (Internal);
            \draw (Internal) -| (A) node[pos=0.7, left] {$l_a$};
            \draw (Internal) -| (B) node[pos=0.7, right] {$l_b$};

            \node[below] at (A) {A};
            \node[below] at (B) {B};
            \node[above right] at (Internal) {P};
        \end{tikzpicture}
    \end{minipage}
    \caption{Geometric diagram of the ancestor state reconstruction algorithm. }
    \label{fig:algorithm-asr}
\end{figure}

In a post-order traversal of the phylogenetic tree, for a selected pair
of sister nodes (sister groups), let $\mathbf{v}_a, \mathbf{v}_b$
be the state vectors of them on a unit hypersphere, and $l_a, l_b$ be their branch
lengths (for leaf nodes) or the equivalent branch lengths (for internal nodes)
on a timetree. Let $\mathbf{v}_p$ be the feature vector of their parent node
(most recent common ancestor), it must be on the two-dimensional Euclidean plane spanned
by $\mathbf{v}_a$ and $\mathbf{v}_b$. The ancestral state and the contrast is computed
via the following steps \cref{fig:algorithm-asr}:

\begin{enumerate}
    \item We firstly calculate their geodesic distance, which is equivalent to their included angle
            for $L_2$ normalised vectors:
    \begin{equation}
        \theta = \arccos{(\mathbf{v}_a\cdot\mathbf{v}_b)}
    \end{equation}

    \item The parent state $\mathbf{v}_p$ is inferred by interpolating along the geodesic arc:
    \begin{equation}
        \mathbf{v}_p = \frac{\sin((1-t)\theta)}{\sin\theta}\mathbf{v}_a + \frac{\sin(t\theta)}{\sin\theta}\mathbf{v}_b
    \end{equation}
    where $t = l_b / (l_a + l_b)$.

    \item To adjust for the curvature of the state space, the contrast variance is scaled by the ratio of the squared arc length to the squared chord length:
    \begin{equation}
        \text{Contrast}^2 = \underbrace{\frac{(\mathbf{v}_a - \mathbf{v}_b)^{\circ 2}}{l_a + l_b}}_{\text{Euclidean Variance}} \times \underbrace{\frac{\theta^2}{||\mathbf{v}_a - \mathbf{v}_b||^2}}_{\text{Correction Factor}}
    \end{equation}
    where $\circ$ means Hadamard power (element-wise power). In standard Phylogenetic Comparative Methods (PCMs), the phenotypic distance between two states is measured as the linear chord length $||v_a - v_b||$. However, on a Riemannian manifold, the true evolutionary path follows the geodesic curve, with distance measured by the arc length $\theta$. For closely related species, the chord length approximates the arc length. However, for highly divergent taxa across deep evolutionary time, the linear chord length severely underestimates the true evolutionary distance. 

    \item When the traversal is completed, the estimated evolutionary rate of the whole tree is:
    \begin{equation}
        \hat{\sigma}^2 = \frac{\sum\text{Contrast}^2}{N-1}
    \end{equation}
    where $N$ is the numbers of leaf nodes (species).
\end{enumerate}

Once the whole ASR process is finished, the spherical variance for every time slice (1 ma) is calculated
on the whole timetree, which represents the phenotypic disparity of all living branches at the specific time.
Then, we conduct 100 times of null simulations based on Brownian motion model in preorder traversals.
For every internal node, the feature vectors of its children are calculated
according to the following algorithm (\cref{fig:algorithm-bm}):

\begin{figure}[htbp]
    \centering
    \begin{tikzpicture}[>=Stealth, thick, scale=0.7]

        % --- CONFIGURATION ---
        \def\R{6}          % Radius of the sphere (length of vp)
        \def\angTheta{45}  % Angle theta
        \def\angAlpha{65}  % Angle of vx relative to vertical
        \def\lenVx{4.5}    % Length of vx vector

        % Coordinates
        \coordinate (Origin) at (0,0);           % The "North Pole" (Tangent point)
        \coordinate (Center) at (0, -\R);        % The Center of the sphere

        % --- AXES ---
        % Vertical dashed axis (upwards)
        \draw[dashed, thin] (0, 0) -- (0, 2.5);

        % Horizontal dashed axis (Tangent Space)
        \draw[dashed, thin] (-2, 0) -- (7, 0) node[right] {tangent space};

        % --- TOP PART: PROJECTION ---

        % Vector vx
        \coordinate (Vx) at (\angAlpha:\lenVx); % Defined by angle alpha
        % Note: In the image, vx is (x,y). Let's approximate the slope visually.
        \coordinate (Vx_End) at (5, 2);
        \draw[->, thin] (Origin) -- (Vx_End) node[midway, above left] {$\mathrm{v_x}$};

        % Vector vt (Projection on horizontal)
        \coordinate (Vt_End) at (5, 0);
        \draw[->, thin] (Origin) -- (Vt_End) node[midway, above] {$\mathrm{v_t}$};

        % Projection line 'a'
        \draw[dashed, thin] (Vx_End) -- (Vt_End) node[midway, right] {$\mathrm{a}$};

        % Angle Alpha
        \coordinate (Y_Axis_Point) at (0, 2);
        \pic [draw, pic text=$\alpha$, angle radius=0.8cm, angle eccentricity=1.3] {angle = Vx_End--Origin--Y_Axis_Point};

        % --- BOTTOM PART: GEODESIC UPDATE ---

        % Calculate the "Child" point on the sphere surface
        % x = R * sin(theta)
        % y = -R + R * cos(theta) (relative to origin)
        % But we draw from Center (0, -R).
        % Target point S relative to Center is (R sin theta, R cos theta).

        \coordinate (S) at ({\R*sin(\angTheta)}, {-\R + \R*cos(\angTheta)});
        \coordinate (M) at (0, {-\R + \R*cos(\angTheta)}); % The corner of the triangle

        % 1. Vertical Component (vp * cos theta)
        % Draws from Center to M
        \draw[->, thin] (Center) -- (M) node[midway, left=5pt] {$\mathrm{v_p} \cdot \cos\theta$};

        % 2. Horizontal Component (vt/|vt| * sin theta)
        % Draws from M to S
        \draw[->, thin] (M) -- (S) node[midway, above] {$\frac{\mathrm{v_t}}{\|\mathrm{v_t}\|} \sin\theta$};

        % 3. Result Vector (vc)
        % Draws from Center to S
        \draw[->, thin] (Center) -- (S) node[midway, below right] {$\mathrm{v_c}$};

        % 4. Parent Vector (vp) label logic
        % The line goes from Center to Origin.
        % We need an arrow head at the Origin.
        \draw[->, thin] (Center) -- (Origin); % completing the vertical line
        \node[above=40pt, left=5pt] at (0, -2.5) {$\mathrm{v_p}$}; % Label roughly in the middle of the top section

        % 5. Angle Theta
        \pic [draw, pic text=$\theta$, angle radius=1.0cm, angle eccentricity=1.3] {angle = S--Center--M};

        % 6. The Geodesic Arc
        % Arcs from Origin (top) down to S.
        % To draw an arc in TikZ, we specify start angle, end angle, and radius.
        % Start: 90 degrees (relative to Center). End: 90 - theta.
        \draw[thick] (0,0) arc (90:{90-\angTheta}:\R);
    \end{tikzpicture}
    \caption{Geometric diagram of the brownian motion simulation algorithm. }
    \label{fig:algorithm-bm}
\end{figure}
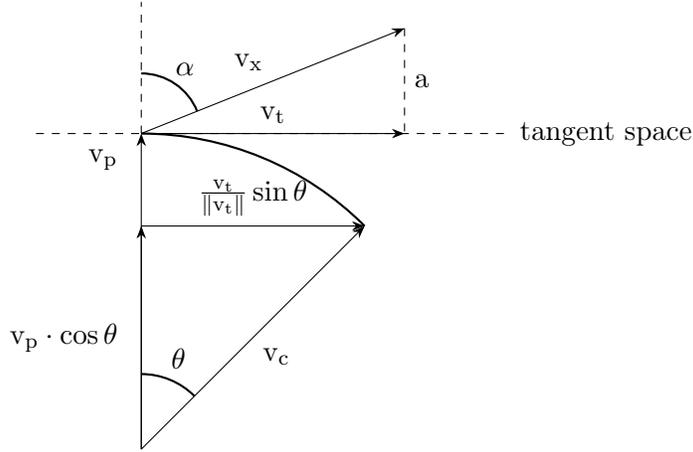

\begin{enumerate}
    \item Create a Gaussian noise $\mathbf{v}_{x}$ in the Euclidean space following this multivariate normal distribution:
    \begin{equation}
        \mathbf{v}_{x} \sim \mathcal{N}(\mathbf{0}, \hat{\sigma}^2 t \mathbf{I}_D)
    \end{equation}
    where $t$ is the evolution time (branch length) of the child node.
    \item Project $\mathbf{v}_{x}$ onto the tangent space of the hypersphere:
    \begin{equation}
        a = \mathbf{v}_{p} \cdot \mathbf{v}_{x}
    \end{equation}
    \begin{equation}
        \mathbf{v}_{t} = \mathbf{v}_{x} - a \mathbf{v}_{p}
    \end{equation}
    \item Move on the hypersphere following a great circle, the direction and geodesic distance is same as the direction and the norm of $\mathbf{v}_{t}$:
    \begin{equation}
        \theta = ||\mathbf{v}_{t}||
    \end{equation}
    \begin{equation}
        \mathbf{v}_{c} = \cos{\theta}~\mathbf{v}_{p} + \frac{\sin{\theta}}{||\mathbf{v}_{t}||} \mathbf{v}_{t}
    \end{equation}
    and $\mathbf{v}_{c}$ is the simulational feature vector of the child species.
\end{enumerate}

When the whole traversal is finished, the ASR is conducted based on the simulational
feature vectors of all modern species (leaf nodes) following the aforementioned algorithm,
as well as the spherical variance for every time slice. Finally, the empirical disparity
through time result and the mean disparity through time of all 100 times of null simulations
are visualised and compared. 

\section{Results}\label{sec:results}

\subsection{Model and its basic characteristics}\label{subsec:r-model}

\begin{figure}[htbp]
    \centering
    \includegraphics[width=\textwidth]{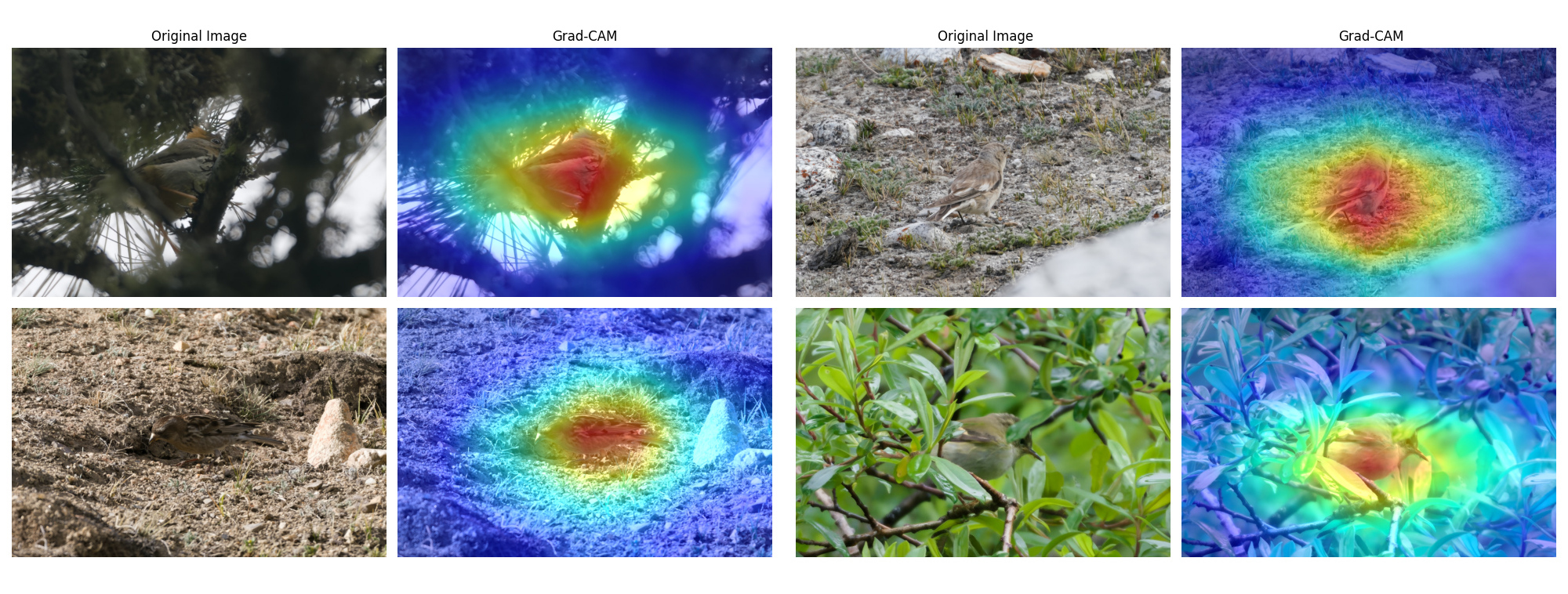}
    \caption{Grad-CAM reveals that the network's attention is consistently
    focused on birds, ignoring backgrounds and occlusions.}
    \label{fig:gradcam}
\end{figure}

The model achieved 90.8\% accuracy on the training set and 87.6\% accuracy
on the validation set. Grad-CAM reveals that the network's attention is focused
on birds, effectively ignoring complex backgrounds and occlusions. The attention
maps cover the entire bird, indicating that the extracted phenotypic vectors
can be regarded as representations of the shape, plumage, and colour of species (\cref{fig:gradcam}).

\begin{figure}[htbp]
    \centering
    \includegraphics[width=\textwidth]{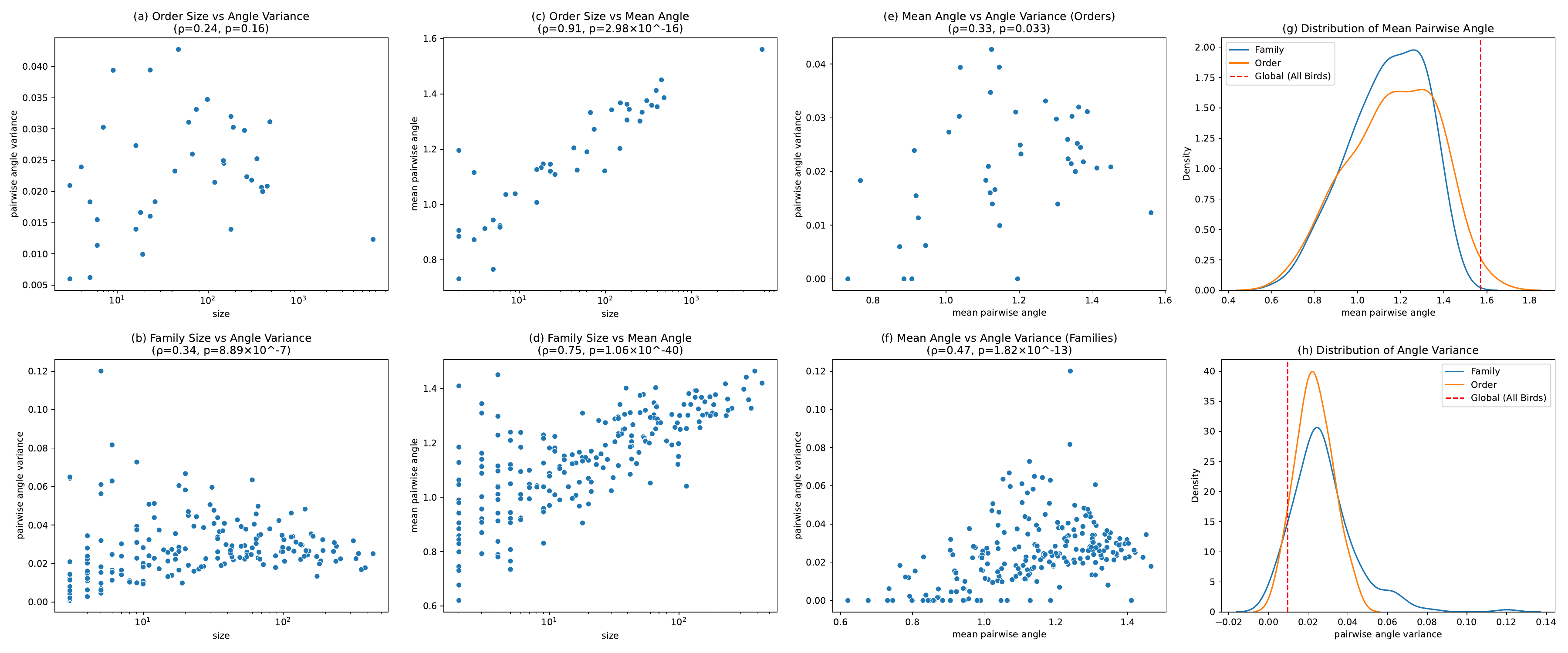}
    \caption{(a-f) The relationship between taxa size and mean pairwise angle,
        taxa size and pairwise angle variance, as well as mean pairwise angle and pairwise
        angle variance at both order and family levels. Spearman's rank coefficients ($\rho$)
        and p-values are listed for each figure; (g-h) the distribution of mean pairwise angle
        and pairwise angle variance of families and orders, the mean values of all birds are
        marked as the red dash line.}
    \label{fig:angles}
\end{figure}

The global mean pairwise angle across all birds is approximately
1.57066 radians ($\approx \frac{\pi}{2}$), with a low variance
($\approx 0.00968$). In contrast, the intra-taxa distributions (families and orders) exhibit
significantly lower mean angles ($\approx1.2$) but higher variances
(peaking at $\approx0.02\textminus0.03$) (\cref{fig:angles}).

A strong positive correlation exists between taxa size
and mean pairwise angle (Orders: $\rho=0.91$; Families: $\rho=0.75$).
Mean pairwise angle shows weak correlation with angle variance
(Orders: $\rho=0.33$; Families: $\rho=0.47$).
Conversely, taxa size shows weak or no correlation with angle variance
(Orders: $\rho=0.24,p>0.05$; Families: $\rho=0.34$).

\subsection{Similarity clustering}\label{subsec:r-similarity-clustering}

The clustering process results in a comprehensive hierarchical clustering output.
The agglomerative clustering technique applied to the cosine similarity
measures of the weight vectors yielded a dendrogram that illustrates the
relationships between the different avian species based on their
morphological features learned by the ResNet34 model.

\begin{figure}[htbp]
  \centering
  \begin{minipage}{0.45\textwidth}
    \centering
    \includegraphics[width=1.05\textwidth]{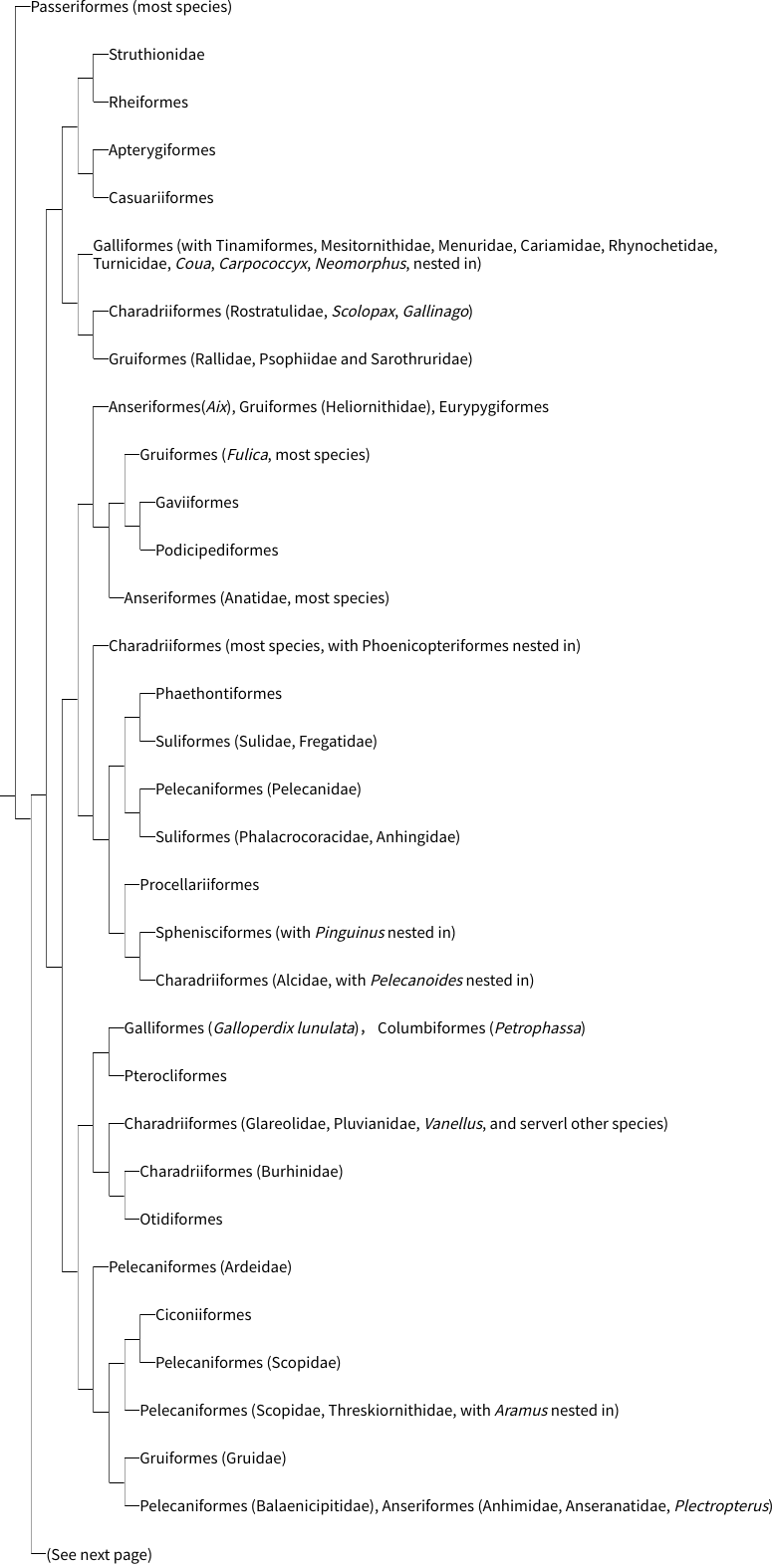}
  \end{minipage}
  \hfill
  \begin{minipage}{0.45\textwidth}
    \centering
    \includegraphics[width=\textwidth]{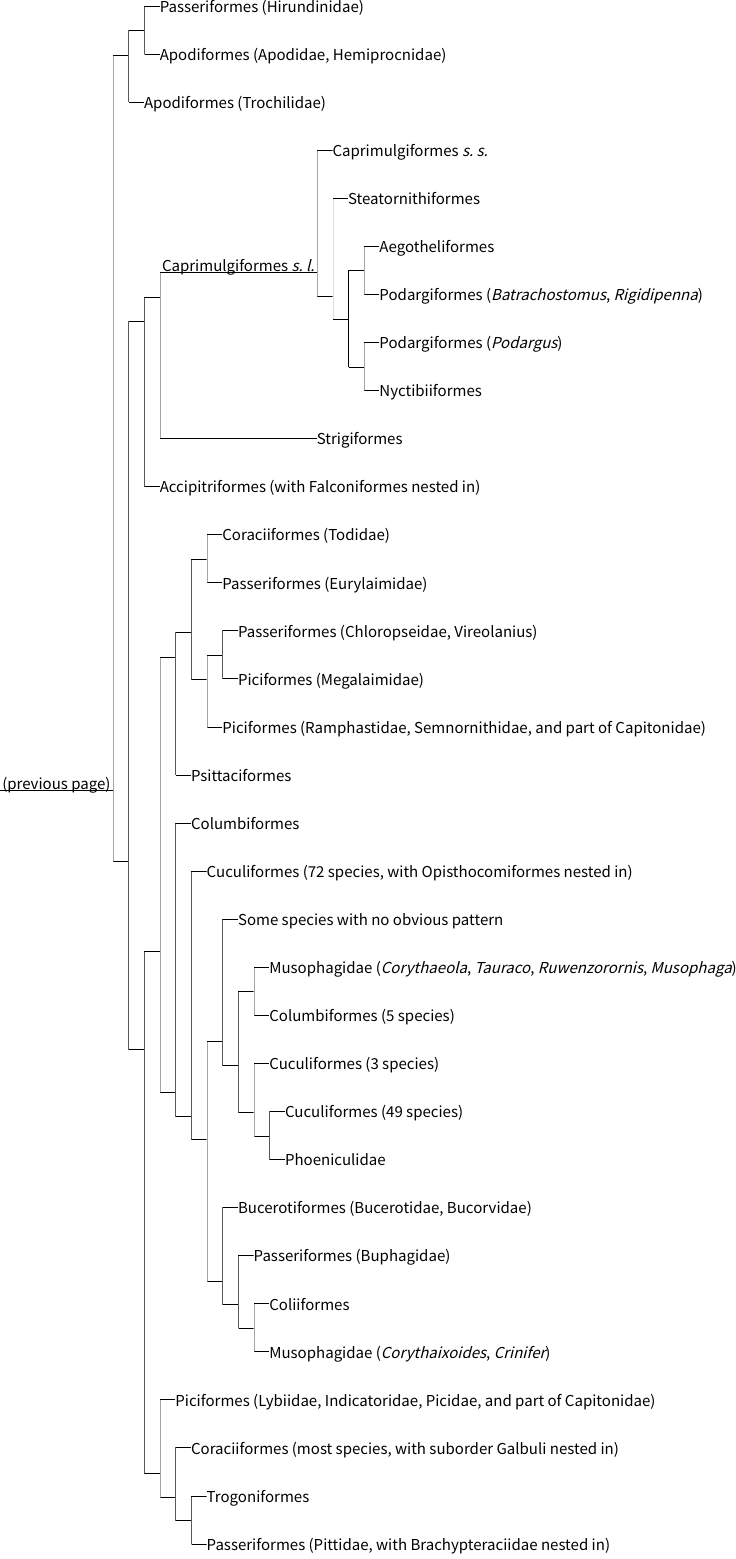}
  \end{minipage}
  \caption{The unrooted clustering result with high-purity branches collapsed.}
\label{fig:cluster}
\end{figure}

In the taxonomic consistence analysis, we identified a total of 391
branches with high taxonomic consistency at the family level and 94
branches at the order level. Additionally, we found 474 family-level
outlier species and 533 order-level outlier species (\cref{fig:cluster}).

\subsection{Disparity analysis}\label{subsec:r-disparity-analysis}

Regarding the relationship between order-level disparity and diversity, the
Spearman's coefficient is at 0.966, with a p-value of $1.45\times10^{-24}$.
For families, the Spearman's coefficient was 0.908,
with a p-value of $3.56\times10^{-83}$.

\begin{figure}[htbp]
  \centering
  \begin{subfigure}{0.48\textwidth}
    \centering
    \includegraphics[width=\textwidth]{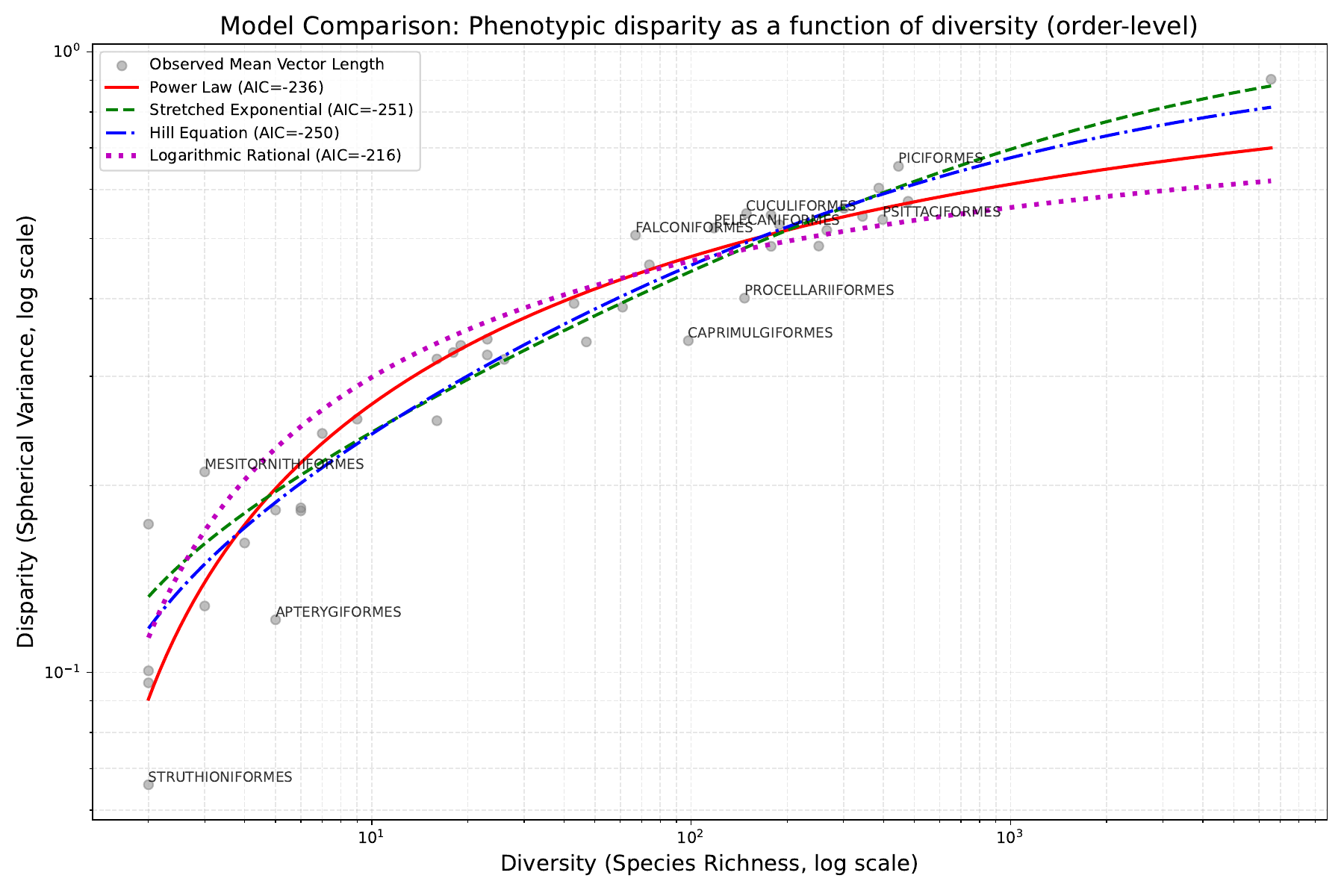}
    \caption{}
    \label{subfig:models_order}
  \end{subfigure}
  \hfill
  \begin{subfigure}{0.48\textwidth}
    \centering
    \includegraphics[width=\textwidth]{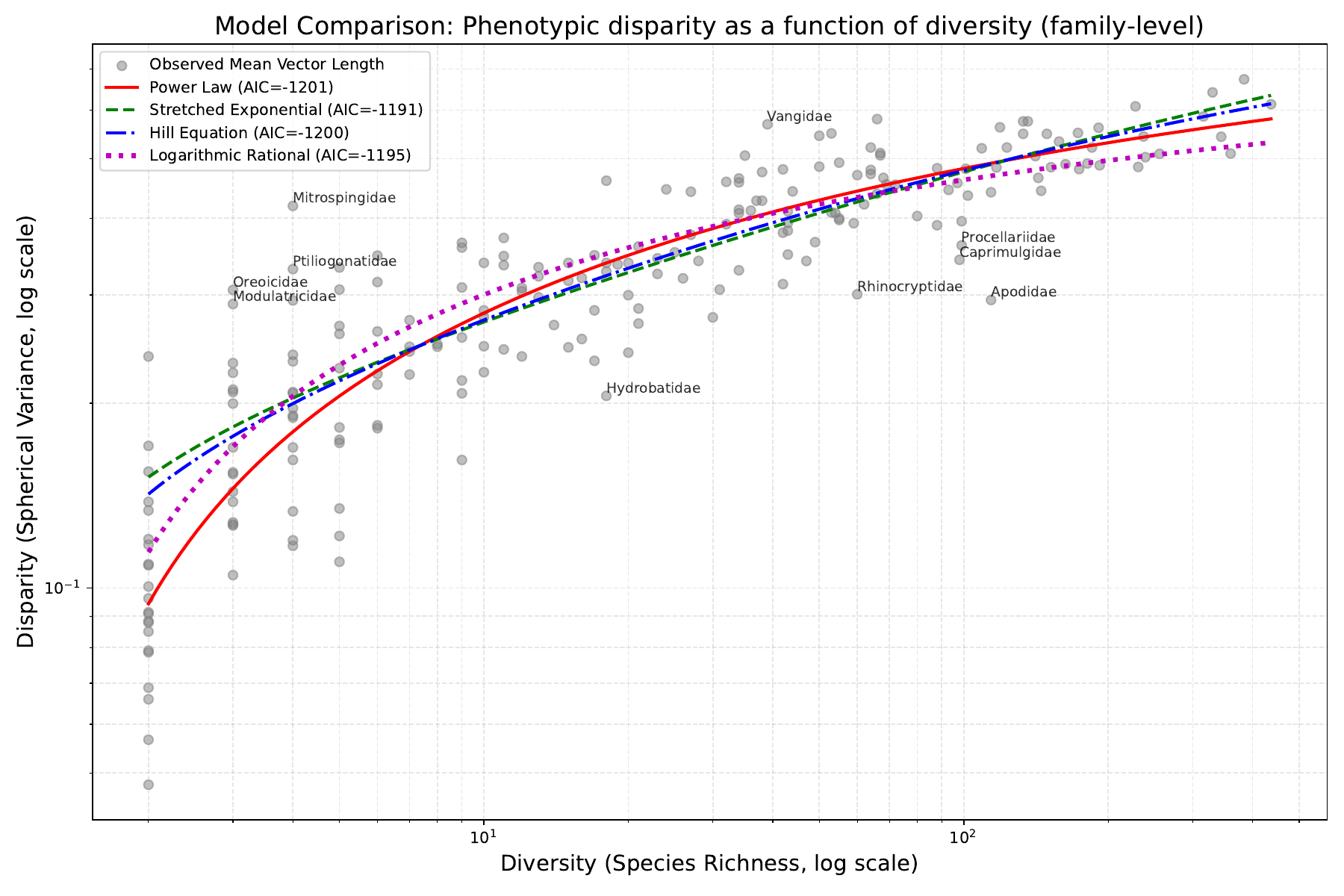}
    \caption{}
    \label{subfig:models_family}
  \end{subfigure}
  \caption{Comparison of models of the phenotypic disparity of taxa (orders or famalies) as a
  function of diversity. Showing scatter plot of all taxa, graphs of four models, names of
  the five taxa with the highest and the five with the lowest phenotypic disparity.}
  \label{fig:models_comparison}
\end{figure}

The stretched exponential model receives the lowest AIC (-251.33) at the order
level, followed by the Hill equation (-250.02), without significant difference
($\Delta AIC = 1.31$). The models are respectively $f(x) = 1 - \exp(-0.1420(x - 1)^{0.3081})$
and $f(x) = \frac{(x - 1)^{0.3973}}{7.4957 + (x - 1)^{0.3973}}$.
Orders with the highest disparity residuals are Falconiformes, Cuculiformes,
Pelecaniformes, Mesitornithiformes (highest first), and Piciformes.
Caprimulgiformes, Procellariiformes, Apterygiformes, Struthioniformes, and Psittaciformes
(lowest first, the same applies below) show the least disaprity residuals (\cref{subfig:models_order}).

For families, the power law receives the lowest AIC (-1201.08), followed by the
Hill equation (-1199.76), without significant difference
($\Delta AIC = 1.32$). The models are respectively $f(x) = 1 - x^{-0.1429}$
and $f(x) = \frac{(x - 1)^{0.3723}}{6.0303 + (x - 1)^{0.3723}}$.
Mitrospingidae, Vangidae, Oreoicidae, Ptiliogonatidae, and Modulatricidae show the
highest disparity residuals, and Apodidae, Rhinocryptidae, Caprimulgidae, Hydrobatidae,
and Procellariidae show the lowest (\cref{subfig:models_family}).

\subsection{Phylogenetic signal}\label{subsec:r-phylogenetic-signal}

The visual morphospace exhibit a highly significant phylogenetic signal ($p = 0.001$, 999 permutations). 
It yields an overall $K_{mult}$ of 0.4150. 
The first phylogenetically aligned component (PAC1) alone exhibited extreme univariate Blomberg's K ($K_{uni} = 2.0775, p = 0.001$).
The first 23 PACs show a $K_{mult}$ of 1.0129, and the value crosses the threshold of 1.0 at PAC24.
In the permutation test, the maximum permuted $K_{mult}$ is 0.3627, and the maximum permuted $K_{uni}$ of PAC1 is 1.3290.

\begin{figure}[htbp]
    \centering
    \includegraphics[width=\textwidth]{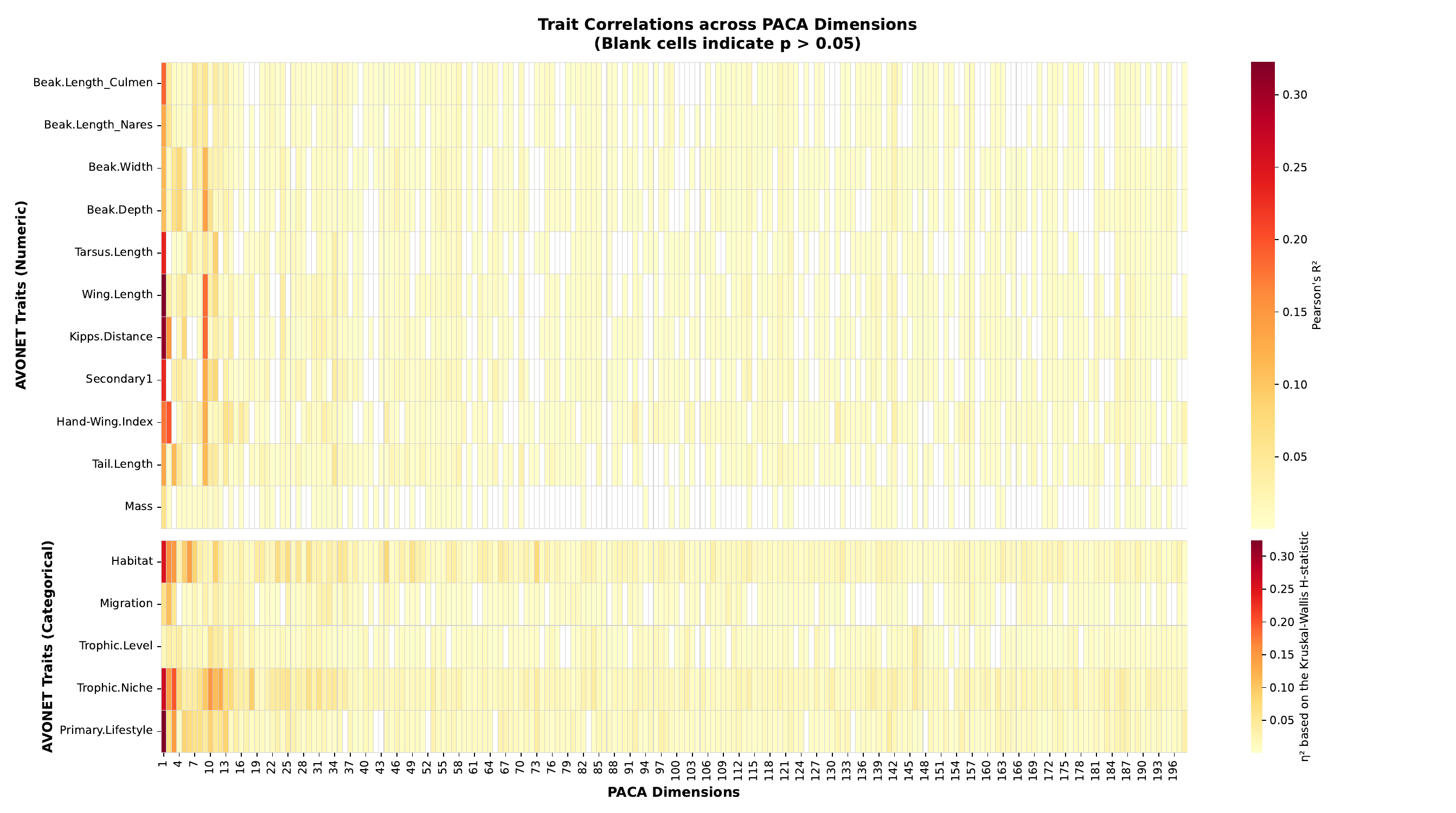}
    \caption{Alignment between the CNN's embedding space with biological traits, represented by the proportion of explained variance. The horizontal axis represents the components of PACA (Phylogenetically Aligned Component Analysis), sorted by phylogenetic signal strength. The vertical axis represents bird traits in AVONET \parencite{tobias2022avonet}. Numerical traits are quantified by Pearson's $R^2$, and categorical traits are quantified by the $\eta^2$ based on Kruskal-Wallis H-statistics. Blank cells indicate statistical insignificance of the correlation ($p > 0.05$). Trait abbreviations are defined as same as \textcite{tobias2022avonet}.}
    \label{fig:paca-avonet}
\end{figure}

\begin{table}[htbp]
  \centering
  \caption{The PACA axis explained maximum variance of each AVONET trait. Trait abbreviations are defined as same as \textcite{tobias2022avonet}. }
  \label{tab:paca-avonet}
  \begin{tabular}{lll}
    \toprule         
    Trait & Max explained variances  &  PACA axes \\ 
    \midrule                   
    Beak.Length\_Culmen & 0.186940 &            1 \\ 
    Beak.Length\_Nares  & 0.131211 &            1 \\ 
    Beak.Width          & 0.114404 &            9 \\ 
    Beak.Depth          & 0.139089 &            9 \\ 
    Tarsus.Length       & 0.235985 &            1 \\ 
    Wing.Length         & 0.322378 &            1 \\ 
    Kipps.Distance      & 0.309666 &            1 \\ 
    Secondary1          & 0.231521 &            1 \\ 
    Hand-Wing.Index     & 0.197398 &            2 \\ 
    Tail.Length         & 0.131632 &            1 \\ 
    Mass                & 0.058428 &            1 \\ 
    Habitat             & 0.251840 &            1 \\ 
    Migration           & 0.109775 &            2 \\ 
    Trophic.Level       & 0.061948 &           10 \\ 
    Trophic.Niche       & 0.265584 &            1 \\ 
    Primary.Lifestyle   & 0.323673 &            1 \\ 
    \bottomrule                 
  \end{tabular}
\end{table}

As shown in \cref{fig:paca-avonet,tab:paca-avonet}, early (approximately first 10) PACA axes (highest phylogenetic signal) show the strongest explained variance (darkest red/orange) with most traits. Correlations decrease quickly as PACA dimension index increases, while do not go to zero entirely. The vast majority of traits peak at PACA dimension 1, which is the single axis of maximum phylogenetic signal. Mass, Trophic.Level, and Migration show low max explained variance.

\subsection{Disparity through time}\label{subsec:r-disparity-through-time}

\begin{figure}[htbp]
    \centering
    \includegraphics[width=0.65\textwidth]{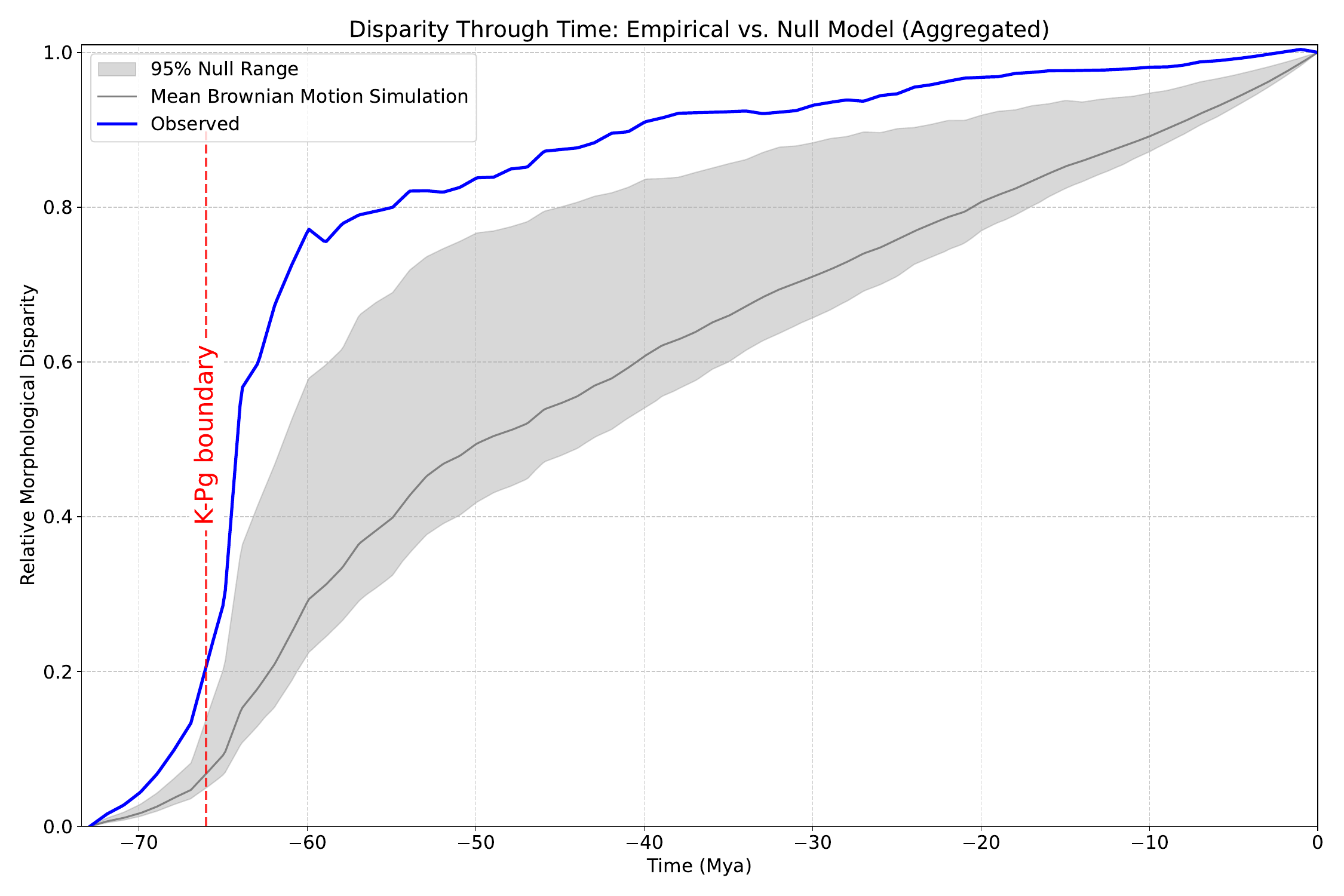}
    \caption{Relative disparity-through-time (DTT) plot for avian visual morphospace.
    The solid blue line represents the empirical disparity, while the solid grey line
    indicates the mean expectation under a multivariate Brownian Motion (BM) null model.
    Both empirical and simulational data are normalised relative to the disparity of
    the extant time (0 Mya). The K-Pg boundary (66 Mya) is marked as the red dash line.}
    \label{fig:dtt}
\end{figure}

The disparity-through-time (DTT) analysis reveals an early-burst pattern for the evolution of
avian visual morphospace. As shown in \cref{fig:dtt}, the empirical data of relative disparity
deviates extremely from the expectation of Brownian Motion null simulation.

Following the K-Pg mass extinction ($\sim 66$ Mya), the empirical disparity immediately increases
at an intense rate. This curve is consistently higher than the mean null expectation throughout
the Paleogene and Neogene, resulting in a positive Morphological Disparity Index (MDI).

Crucially, the empirical curve achieves approximately 50\% of the modern disparity shortly after
the K-Pg mass extinction, followed by a relative deceleration towards the present. Conversely,
the BM model requires significantly more time to reach the same disparity.

\section{Discussion}\label{sec:discussion}

\subsection{Geometry of the morphospace}\label{subsec:geometry-of-the-morphospace}

The geometric analysis of the feature space reveals a distinctive interplay
between high-dimensional properties and evolutionary constraints.
The mean pairwise angle of $\frac{\pi}{2}$ aligns with the
property of high-dimensional spaces where random vectors tend to
be orthogonal and equidistant (\cref{fig:angles}).
Intra-taxa pairwise angles, although varied across taxa, are
smaller than the global mean value.
Since it is trained solely on species classification (flat labels),
a hierarchical taxonomic structure (Order/Family) automatically emerges in the high-dimensional feature space.
This demonstrates that feature vectors are semantically clustered,
a biological taxa occupying specific sub-manifolds and therefore exhibiting smaller
mean pairwise angle. This indicates that biological relationships are strongly encoded
in morphological features that models have to learn this structure when
searching for the optimal classification structure.

In the raw result of the clustering, many extinct species clustered together,
possibly due to their representation via skeletal images, artistic reconstructions, or
other non-biological patterns, being fundamentally different from extant species.
Furthermore, some of the recently described species or newly separated cryptic species grouped
together, being indistinguishable, which might reflect insufficient image data, leading to
underfitting of the model. All extinct species and some of the newly described or
newly split species, totally 249 species, were removed in subsequent analyses.

Although the clustering result is not perfectly align with current taxonomy,
these inconsistencies exhibits some historically noted morphological similarity.

In the Palaeognathae, four orders (Struthioniformes, Rheiformes, Casuariiformes, Apterygiformes)
of flightless birds form a distinct cluster, being driven by their unique body plans. In contrast,
Tinamiformes represents a notable exception, which is deeply within the Galliformes.
This topological incongruence highlights their convergent evolution. Tinamiformes and many Galliformes
have evolved highly convergently, exhibiting compact body shapes for ground-dwelling lifestyles,
and occupy similar niches \parencite{Glenny1946Systematic}. The model correctly identifies
this ecological and morphological similarity (\cref{fig:cluster}).

Furthermore, the inclusion of other visually quail-like terrestrial taxa like
Turnicidae (Buttonquails) \parencite{Rotthowe1998Evidence}, Menuridae \parencite{Winkler2020Lyrebirds},
Neomorphus \parencite{Haffer1977Neomorphus}
within the Galliformes cluster further validates that our model's feature space captures
morphological and ecological convergent adaptation and visual similarity. In the eyes of CNNs,
as long as you scratch for food and run around on the ground like a landfowl,
then you are a "landfowl" (\cref{fig:cluster}).

The clustering of Caprimulgiformes \textit{sensu lato}, Strigiformes, Accipitriformes, and Falconiformes
forms a visually similar group (\cref{fig:cluster}), with two layers of convergent evolution: Strigiformes and Caprimulgiformes \textit{s. l.}
share large eyes and cryptic plumage patterns for their nocturnal habits; and Strigiformes, Accipitriformes,
and Falconiformes share the raptorial facial characteristics like forward-facing eyes and hooked beaks
\parencite{Fidler2004Convergent, Wu2021Genomic}.

Other notable similarity exhibited in the clustering result includes
Hirundinidae and Apodi (Apodidae and Hemiprocnidae) \parencite{Norberg1986Insectivorous},
Sphenisciformes (Penguins), Alcidae and Procellariiformes \parencite{Cody1973Coexistence},
as well as Phaethontiformes, Suliformes, and Pelecanidae \parencite{Hedges1994Molecules}~\ldots~(\cref{fig:cluster})
It's demonstrated that the feature vectors' distribution in morphological space is not
arbitrary, but is constrained by objective phenotypic similarity. Considering that we did not
introduce \textit{a priori} taxonomic knowledge in the training process, it actually reinvented
the classical taxonomy from Linnaeus to Audubon. The model's taxonomy is based on phenotype,
yet with higher throughput of data than classical taxonomy and morphometrics.
This naturally creates a gap compared to molecular phylogenetic trees.
The gap represents the disconnection between visual disparity and genetic distance.

\subsection{The implications of the phylogenetic signals}\label{subsect:phylosignal-discussion}

In the phylogenetic signal analysis, the value $K_{mult} < 1$ indicates that the overall phylogenetic signal is lower than expected under Brownian motion.
The embedding space has moderate phylogenetic signal, which is not random with respect to phylogeny.
This confirms that the CNN embedding is phylogenetically structured globally, and the phylogenetic constraints shaped the bird visual morphospace.
While it furthermore indicates that the lineage divergence is also coupled with convergent evolution and plasticity. 

The univariate $K$ value of the first PAC demonstrates that the primary axis of visual variation is strongly constrained by shared ancestry. 
Phylogenetically close species are twice as similar along this axis as would be expected under neutral evolution. 
The empirical value exceeds the permutation maximum by a factor of about 1.56 times, which is a very unlikely result under the null hypothesis. 
This indicates that visual deep learning on natural organisms can recover authentic evolutionary structure, and distinguish between convergent and divergent evolution.

As shown in \cref{fig:paca-avonet,tab:paca-avonet}, the leftmost columns (dimensions 1--10) show the darkest colours (highest $R^2$/$\eta^2$), indicating that the first few PACA components -- those with the strongest phylogenetic signal -- correlate most strongly with nearly all AVONET traits. This means that the CNN has learned features that implicitly reflect evolutionary relationships, with the most phylogenetically-structured dimensions encoding the most biologically meaningful information. 

Combined with the aforementioned qualitative analysis of visual phenotypic similarity, it is evident that the CNN embedding space is not only mathematically structured, but also biologically meaningful. The model has learned to encode visual features that are both phylogenetically informative and ecologically relevant, capturing the complex interplay between morphology, function, and evolutionary history. 

Furthermore, the model has captured both homologous and convergent traits and disentangled them from each other. 
If homologous and convergent traits were entangled, we would expect K values to be more uniform across all axes, no significant gradients, as we see in the permutation test. Instead, the first few PACs show significantly stronger phylogenetic signal, indicating that the model has learned to separate traits that are conserved across lineages from those that have evolved convergently. 

Additionally, if a CNN trained on photographs recovers this structure, it implies that the structure exists in the phenotype themselves. 
In other words, homologous and convergent traits are inherently orthogonal in the natural world. 
This orthogonality is an inherent property of biological visual variation shaped in the evolution. 
The CNN simply discovers this pre-existing orthogonality after being trained on sufficient data without being given biological concepts. 

\subsection{Shape vs Texture}\label{subsec:shape-vs-texture}

Recent critiques in computer vision suggest that CNNs are biased towards
textural features rather than the whole shapes \parencite{Geirhos2018ImagenetTrained}.
However, the aforementioned discussion on the biological semantic of the embedding space implies
that our model mainly focuses on the body plan rather than textures,
challenge this assumption in the context of fine-grained visual classification.

For instance, in the previously discussed phenotypic resemblances, Hirundinidae and Apodidae
shares similar aerodynamic characteristics for the aerial insectivorous niche.
The instance of Sphenisciformes, Alcidae, and Procellariiformes constitutes a more convincing argument.
Penguins (Sphenisciformes) have highly specialised feathers, being extremely dense,
thick, and visually scale-like \parencite{Williams2015Hidden}.
Although \textcite{Williams2015Hidden} demonstrates that
emperor penguins (\textit{Aptenodytes forsteri}) do have filoplumes and plumules,
they are highly likely to be invisible in most images.
While auks (Alcidae), albatrosses, petrels and shearwaters (Procellariiformes) have typical and normal feathers.
If a model is textures biased, the three orders cannot be placed adjacently in the morphospace.

This hypothesis is further supported by our Grad-CAM analysis (\cref{fig:gradcam}),
which shows activation maps covering the entire avian body rather than specific patches.
This indicates that our model are capable of capturing holistic representations,
mirroring the concept of 'body plan' in biology.

% To test this hypothesis, we used Google Nano Banana (gemini-2.5-flash-image)
% to generate a penguin with typical feathers. Our model successfully classified this
% adversarial example as an Adélie penguin (\textit{Pygoscelis adeliae}) with the highest confidence,
% and the top-3 predictions are all penguins (\cref{fig:penguin}).
% This finding suggests that the feature vectors captured by our deep learning framework is biologically grounded.
% The model has learned knowledge on the macrostructure (body plan) rather than merely textures,
% thereby reinforcing the biological credibility of the subsequent morphological disparity analysis.

% \begin{figure}[htbp]
%     \centering
%     \includegraphics[width=0.35\textwidth]{chicken-plumaged-penguin}
%     \caption{The adversarial example of a chicken-plumaged penguin generated with Google Nano Banana.
%         It was cropped to the bird using Faster R-CNN, and then fed into the model.
%         Results are: Adélie penguin (\textit{Pygoscelis adeliae}) 36.51\%,
%         yellow-eyed penguin (\textit{Megadyptes antipodes}) 11.67\%, and
%         little penguin (\textit{Eudyptula minor}) 8.58\%.}
%     \label{fig:penguin}
% \end{figure}

\subsection{Macroevolution of birds}\label{subsec:macroevolution-of-birds}

The Spearman's rank correlation demonstrates that species richness is the primary driver of morphospace expansion.
The strong positive correlation between taxon size and hyperspherical variance (Orders: $\rho = 0.966$; Families: $\rho = 0.908$)
confirms that as clades diversify, they tend to colonise a larger volume of the visual feature space.
This pattern is consistent with the "niche filling" model \parencite{Simpson1944Tempo},
where speciation events are often associated with the exploration of novel regions
in the phenotypic landscape to minimise competition.
However, this expansion exhibits diminishing returns (\cref{fig:models_comparison}),
suggesting that the marginal gain in disparity decreases as taxa become extremely large.

In contrast, structural heterogeneity appears to be constrained that is largely independent of diversity.
At the order level, angle variance is statistically independent of clade size ($\rho=0.24,p>0.05$),
and at the family level, the correlation is weak ($\rho=0.34$).
This implies that even as a clade radiates swiftly, its internal geometric uniformity
does not change chaotically but remains confined within a stable range.

In the top-5 families with high disparity residuals, four of them are newly recognised based on
molecular phylogeny, indicating that they exhibit greater heterogeneity than long-established taxa,
which is why they remained unrecognised by zoologists for a long time.

Vangidae is a textbook example of the radiation evolution.
Although there aren't many species (39), they have evolved a variety of forms on Madagascar,
ranging from shrike-like and flycatcher-like to finch-like species \parencite{Reddy2012Diversification, Jonsson2012Ecological}.
The model captures this extreme morphological difference, producing a very high positive residual.

Similarly, Cotingidae is given the seventh-highest disparity residual among families.
Their differences in appearance are extremely eye-catching, ranging from the
brightly coloured cocks-of-the-rock (\textit{Rupicola})
to the relatively dull-coloured pihas (\textit{Lipaugus} and \textit{Snowornis}),
with huge differences \parencite{Winkler2020Cotingas}.

The disparity-through-time (DTT) analysis (\cref{fig:dtt}) exhibits a convex shape,
suggesting that morphological disparity increased rapidly after the K-Pg mass extinction.
This furtherly supports \posscite{Simpson1944Tempo} adaptive radiation and the niche filling hypothesis.
After mass extinctions vacated ecological niches,
birds rapidly evolved extreme morphological diversity to occupy these positions.
The surviving avian lineages rapidly radiated to fill the ecological niches,
establishing the major disparate body plans (visual morphotypes) within a short geological window.
The positive Morphological Disparity Index (MDI), evidenced in \cref{fig:dtt},
reinforces that the initial partitioning of the visual morphospace
was significantly faster than neutral drift expectations.

\subsection{Methodological Universality}\label{subsec:methodological-universality}

The most disruptive and valuable advantage of this new methodology is its absolute, mathematical independence from organismal anatomy. Because the high-dimensional feature vectors are strictly mathematically derived from raw visual semantics rather than homologous, human-plotted anatomical coordinates, the framework exhibits universal applicability. Traditional geometric morphometrics cannot compare the evolutionary rate of an insect's structural adaptation against the co-evolutionary rate of the angiosperm flower it pollinates, because no homologous landmarks exist between a plant and an arthropod.

By stark contrast, the proposed deep learning framework can project entirely disparate phylogenetic trees, representing completely different kingdoms of life, into the exact same unified, high-dimensional embedding space. By mathematically calculating the hyperspherical volume occupied by varying phyla, researchers can now directly and seamlessly compare the macroevolutionary disparity of fungi against mammals, or track the co-evolutionary tempo of deeply intertwined but anatomically alien species. This capacity fundamentally alters the scope of comparative biology, decisively upgrading the discipline from narrow, restrictive intra-clade analyses to a unified, scalable quantification of global biodiversity.

\section{Conclusion}\label{sec:conclusion}

This study demonstrates that deep learning models can function as high-throughput instruments
for quantifying morphological disparity. By extracting weights from a ResNet34 model trained
on over 10,000 bird species, we established a high-dimensional morphospace that aligns with
biological taxonomy and encodes evolutionary convergence,
despite the model being trained on flat labels without \textit{a priori} taxonomic knowledge.

Our investigation into the interpretability of these networks challenges the prevailing
view that CNNs rely primarily on local textures. Through the use of adversarial examples
and Grad-CAM analysis, we provide evidence that the model overcomes texture bias
to learn holistic shape representations, effectively capturing the "body plan".

In the context of macroevolution, our analysis reveals that species richness acts as the
primary driver of morphospace expansion, a pattern consistent with the "niche filling" hypothesis.
Furthermore, the disparity-through-time analysis uncovers a visual "early burst"
following the K-Pg extinction, wherein avian lineages rapidly colonised approximately 50\% of
the modern morphological disparity.

This approach serves as a powerful, scalable new method for macroevolution.
By synthesising computer vision with evolutionary biology,
this framework offers a novel perspective on the drivers of biodiversity,
enabling the analysis of morphological evolution at a scale previously unattainable.

\paragraph{Code and data accessibility}
\begin{itemize}
    \item Code for data processing: \url{https://github.com/sun-jiao/osea}
    \item Code for model training: \url{https://github.com/sun-jiao/MetaFGNet}
    \item Code for biological analysis: \url{https://github.com/sun-jiao/osea_morpho_evo}
    \item The model: \url{https://huggingface.co/sunjiao/osea}
\end{itemize}

\paragraph{Conflict of interests}

The author has no conflict of interests.

\paragraph{Author contribution statements}

JS designed the project, programmed the python script and drafted the manuscript.

\paragraph{Acknowledgements}

JS is supported by Chinese Scholarship Council grant 202508420040. 

\printbibliography

@article{adams2014generalized,
  title={A generalized K statistic for estimating phylogenetic signal from shape and other high-dimensional multivariate data},
  author={Adams, Dean C},
  journal={Systematic biology},
  volume={63},
  number={5},
  pages={685--697},
  year={2014},
  publisher={Oxford University Press}
}

@inproceedings{Ansel2024Pytorch,
    author = {Ansel, J. and Yang, E. and He, H. and Gimelshein, N. and Jain, A. and Voznesensky, M. and Bao, B. and Bell, P. and Berard, D. and Burovski, E. and Chauhan, G. and Chourdia, A. and Constable, W. and Desmaison, A. and DeVito, Z. and Ellison, E. and Feng, W. and Gong, J. and Gschwind, M. and others},
    title = {PyTorch 2: Faster Machine Learning Through Dynamic Python Bytecode Transformation and Graph Compilation},
    booktitle = {Proceedings of the 29th ACM International Conference on Architectural Support for Programming Languages and Operating Systems, Volume 2},
    year = {2024},
    pages = {929--947},
    doi = {10.1145/3620665.3640366}
}

@article{Cody1973Coexistence,
    author = {Cody, Martin L.},
    title = {Coexistence, Coevoluation and Convergent Evolution in Seabird Communities},
    journal = {Ecology},
    volume = {54},
    number = {1},
    pages = {31-44},
    doi = {10.2307/1934372},
    url = {https://esajournals.onlinelibrary.wiley.com/doi/abs/10.2307/1934372},
    eprint = {https://esajournals.onlinelibrary.wiley.com/doi/pdf/10.2307/1934372},
    year = {1973}
}

@article{collyer2021phylogenetically,
  title={Phylogenetically aligned component analysis},
  author={Collyer, Michael L and Adams, Dean},
  year={2021},
  publisher={Wiley on behalf of the British Ecological Society}
}

@article{daboul2018procrustes,
  title={Procrustes-based geometric morphometrics on MRI images: An example of inter-operator bias in 3D landmarks and its impact on big datasets},
  author={Daboul, Amro and Ivanovska, Tatyana and B{\"u}low, Robin and Biffar, Reiner and Cardini, Andrea},
  journal={PloS one},
  volume={13},
  number={5},
  pages={e0197675},
  year={2018},
  publisher={Public Library of Science San Francisco, CA USA}
}

@article{Farid2021Overview,
    author = {Farid, Hany},
    title = {An Overview of Perceptual Hashing},
    journal = {Journal of Online Trust and Safety},
    volume = {1},
    number = {1},
    year = {2021},
    doi = {10.54501/jots.v1i1.24}
}

@article{Felsenstein1985PIC,
    author = {Joseph Felsenstein},
    journal = {The American Naturalist},
    number = {1},
    pages = {1--15},
    publisher = {[The University of Chicago Press, The American Society of Naturalists]},
    title = {Phylogenies and the Comparative Method},
    doi = {10.1086/284325},
    volume = {125},
    year = {1985}
}

@article{Fidler2004Convergent,
    title = {Convergent evolution of strigiform and caprimulgiform dark-activity is supported by phylogenetic analysis using the arylalkylamine N-acetyltransferase (Aanat) gene},
    author = {Fidler, Andrew E and Kuhn, Sylvia and Gwinner, Eberhard},
    journal = {Molecular Phylogenetics and Evolution},
    volume = {33},
    number = {3},
    pages = {908--921},
    year = {2004},
    publisher = {Elsevier},
    doi = {10.1016/j.ympev.2004.08.015},
}

@inproceedings{Geirhos2018ImagenetTrained,
    title = {{ImageNet}-trained {CNN}s are biased towards texture; increasing shape bias improves accuracy and robustness.},
    author = {Robert Geirhos and Patricia Rubisch and Claudio Michaelis and Matthias Bethge and Felix A. Wichmann and Wieland Brendel},
    booktitle = {International Conference on Learning Representations},
    year = {2019},
    url = {https://openreview.net/forum?id=Bygh9j09KX},
    doi = {10.48550/arXiv.1811.12231},
}

@misc{Gill2021IOC,
    author = {Gill, F. and Donsker, D. and Rasmussen, P.},
    title = {{IOC} world bird list 10.1},
    publisher = {International Ornithologists' Union},
    year = {2021},
    doi = {10.14344/IOC.ML.10.1}
}

@misc{Gill2025IOC,
    author = {Gill, F. and Donsker, D. and Rasmussen, P.},
    title = {{IOC} world bird list 15.1},
    publisher = {International Ornithologists' Union},
    year = {2025},
    url = {https://www.worldbirdnames.org/new/ioc-lists/master-list-2/}
}

@article{Glenny1946Systematic,
    title = {A Systematic Study of the Main Arteries in the Region of the Heart -- {Aves XI: Tinamiformes} -- with some notes on their apparent relationship with the {Galliformes}},
    author = {Glenny, Fred H},
    journal = {Canadian Journal of Research},
    volume = {24},
    number = {2},
    pages = {31--38},
    year = {1946},
    publisher = {NRC Research Press Ottawa, Canada},
    doi = {10.1139/cjr46d-004},
}

@software{Gommers2025Scipy,
    author = {Gommers, R. and Virtanen, P. and Haberland, M. and Burovski, E. and Reddy, T. and Weckesser, W. and Oliphant, T. E. and Cournapeau, D. and Nelson, A. and alexbrc and Roy, P. and Peterson, P. and Polat, I. and Wilson, J. and endolith and Mayorov, N. and van der Walt, S. and Colley, L. and Brett, M. and others},
    title = {scipy/scipy: SciPy 1.15.0},
    publisher = {Zenodo},
    year = {2025},
    doi = {10.5281/zenodo.14593523}
}

@article{Guillerme2020Disparities,
    author = {Guillerme, T. and Cooper, N. and Brusatte, S. L. and Davis, K. E. and Jackson, A. L. and Gerber, S. and Goswami, A. and Healy, K. and Hopkins, M. J. and Jones, M. E. H. and Lloyd, G. T. and O'Reilly, J. E. and Pate, A. and Puttick, M. N. and Rayfield, E. J. and Saupe, E. E. and Sherratt, E. and Slater, G. J. and Weisbecker, V. and others},
    title = {Disparities in the analysis of morphological disparity},
    journal = {Biology Letters},
    volume = {16},
    number = {7},
    pages = {20200199},
    year = {2020},
    doi = {10.1098/rsbl.2020.0199}
}

@article{Haffer1977Neomorphus,
    title = {A Systematic Review of the Neotropical Ground Cuckoos ({Aves}, {Neomorphus})},
    journal = {Bonner zoologische Beiträge : Herausgeber: Zoologisches Forschungsinstitut und Museum Alexander Koenig, Bonn},
    volume = {28},
    copyright = {In Copyright. Digitized with the permission of the rights holder.},
    url = {https://www.biodiversitylibrary.org/part/119218},
    publisher = {Bonn, Das Forschungsinstitut, },
    author = {Haffer, Jürgen},
    year = {1977},
    pages = {48--76},
}

@inproceedings{He2016ResNet,
    author = {He, Kaiming and Zhang, Xiangyu and Ren, Shaoqing and Sun, Jian},
    title = {Deep Residual Learning for Image Recognition},
    booktitle = {2016 IEEE Conference on Computer Vision and Pattern Recognition (CVPR)},
    pages = {770--778},
    year = {2016},
    doi = {10.1109/CVPR.2016.90}
}

@article{Hedges1994Molecules,
    title = {Molecules vs. morphology in avian evolution: the case of the" pelecaniform" birds.},
    author = {Hedges, S Blair and Sibley, Charles G},
    journal = {Proceedings of the National Academy of Sciences},
    volume = {91},
    number = {21},
    pages = {9861--9865},
    year = {1994},
    doi = {10.1073/pnas.91.21.9861},
}

@article{Huerta2016ETE,
    author = {Huerta-Cepas, J. and Serra, F. and Bork, P.},
    title = {ETE 3: Reconstruction, Analysis, and Visualization of Phylogenomic Data},
    journal = {Molecular Biology and Evolution},
    volume = {33},
    number = {6},
    pages = {1635--1638},
    year = {2016},
    doi = {10.1093/molbev/msw046}
}

@article{Jonsson2012Ecological,
    title = {Ecological and evolutionary determinants for the adaptive radiation of the Madagascan vangas},
    author = {J{\o}nsson, Knud A and Fabre, Pierre-Henri and Fritz, Susanne A and Etienne, Rampal S and Ricklefs, Robert E and J{\o}rgensen, Tobias B and Fjelds{\aa}, Jon and Rahbek, Carsten and Ericson, Per GP and Woog, Friederike and others},
    journal = {Proceedings of the National Academy of Sciences},
    volume = {109},
    number = {17},
    pages = {6620--6625},
    year = {2012},
    publisher = {National Academy of Sciences},
    doi = {10.1073/pnas.1115835109}
}

@article{LeCun1989Backpropagation,
    author = {LeCun, Y. and Boser, B. and Denker, J. S. and Henderson, D. and Howard, R. E. and Hubbard, W. and Jackel, L. D.},
    title = {Backpropagation Applied to Handwritten Zip Code Recognition},
    journal = {Neural Computation},
    volume = {1},
    number = {4},
    pages = {541--551},
    year = {1989},
    doi = {10.1162/neco.1989.1.4.541}
}

@online{Mei2020Dongniao,
    author = {Mei, J. and Dong, H.},
    title = {The {DongNiao} International Birds 10000 Dataset},
    eprint = {2010.06454},
    eprinttype = {arXiv},
    version = {2},
    year = {2020},
    doi = {10.48550/arXiv.2010.06454}
}

@article{mulqueeney2025assessing,
  title={Assessing the application of landmark-free morphometrics to macroevolutionary analyses},
  author={Mulqueeney, James M and Ezard, Thomas HG and Goswami, Anjali},
  journal={BMC Ecology and Evolution},
  volume={25},
  number={1},
  pages={38},
  year={2025},
  publisher={Springer}
}

@article{Norberg1986Insectivorous,
    ISSN = {00305693},
    URL = {http://www.jstor.org/stable/3676835},
    author = {Ulla M. Norberg},
    journal = {Ornis Scandinavica (Scandinavian Journal of Ornithology)},
    number = {3},
    pages = {253--260},
    publisher = {[Nordic Society Oikos, Wiley]},
    title = {Evolutionary Convergence in Foraging Niche and Flight Morphology in Insectivorous Aerial-Hawking Birds and Bats},
    urldate = {2026-02-02},
    volume = {17},
    year = {1986},
    doi = {10.2307/3676835},
}

@article{Reddy2012Diversification,
    title = {Diversification and the adaptive radiation of the vangas of Madagascar},
    author = {Reddy, S and Driskell, A and Rabosky, DL and Hackett, Shannon J and Schulenberg, TS},
    journal = {Proceedings of the Royal Society B: Biological Sciences},
    volume = {279},
    number = {1735},
    pages = {2062--2071},
    year = {2012},
    publisher = {The Royal Society},
    doi = {10.1098/rspb.2011.2380},
}

@ARTICLE{Ren2017Faster,
    author = {Ren, Shaoqing and He, Kaiming and Girshick, Ross and Sun, Jian},
    journal = {{IEEE} Transactions on Pattern Analysis and Machine Intelligence},
    title = {{Faster R-CNN}: Towards Real-Time Object Detection with Region Proposal Networks},
    year = {2017},
    volume = {39},
    number = {6},
    pages = {1137-1149},
    doi = {10.1109/TPAMI.2016.2577031},
}

@article{Rotthowe1998Evidence,
    title = {Evidence for a phylogenetic position of button quails ({Turnicidae: Aves}) among the {Gruiformes}},
    author = {Rotthowe, K and Starck, JM},
    journal = {Journal of Zoological Systematics and Evolutionary Research},
    volume = {36},
    number = {1-2},
    pages = {39--51},
    year = {1998},
    publisher = {Wiley Online Library},
    doi = {10.1111/j.1439-0469.1998.tb00776.x},
}

@article{Selvaraju2020Grad,
    title = {{Grad-CAM}: Visual Explanations from Deep Networks via Gradient-Based Localization},
    author = {Selvaraju, Ramprasaath R and Cogswell, Michael and Abhishek, Das and Ramakrishna, Vedantam and Devi, Parikh and Dhruv, Batra},
    journal = {International Journal of Computer Vision},
    volume = {128},
    number = {2},
    pages = {336--359},
    year = {2020},
    publisher = {Springer Nature BV},
    doi = {10.1007/s11263-019-01228-7},
}

@book{Simpson1944Tempo,
    title = {Tempo and mode in evolution},
    author = {Simpson, George Gaylord},
    number = {15},
    year = {1944},
    publisher = {Columbia University Press}
}

@article{Stiller2024Complexity,
    title = {Complexity of avian evolution revealed by family-level genomes},
    author = {Stiller, Josefin and Feng, Shaohong and Chowdhury, Al-Aabid and Rivas-Gonz{\'a}lez, Iker and Duch{\^e}ne, David A and Fang, Qi and Deng, Yuan and Kozlov, Alexey and Stamatakis, Alexandros and Claramunt, Santiago and others},
    journal = {Nature},
    volume = {629},
    number = {8013},
    pages = {851--860},
    year = {2024},
    publisher = {Nature Publishing Group UK London},
    doi = {10.1038/s41586-024-07323-1},
}

@article{tobias2022avonet,
  title={AVONET: morphological, ecological and geographical data for all birds},
  author={Tobias, Joseph A and Sheard, Catherine and Pigot, Alex L and Devenish, Adam JM and Yang, Jingyi and Sayol, Ferran and Neate-Clegg, Montague HC and Alioravainen, Nico and Weeks, Thomas L and Barber, Robert A and others},
  journal={Ecology letters},
  volume={25},
  number={3},
  pages={581--597},
  year={2022},
  publisher={Wiley Online Library}
}

@inproceedings{Wang2017NormFace,
    author = {Wang, Feng and Xiang, Xiang and Cheng, Jian and Yuille, Alan Loddon},
    title = {{NormFace}: $L_2$ Hypersphere Embedding for Face Verification},
    year = {2017},
    isbn = {9781450349062},
    publisher = {Association for Computing Machinery},
    address = {New York, NY, USA},
    doi = {10.1145/3123266.3123359},
    booktitle = {Proceedings of the 25th ACM International Conference on Multimedia},
    pages = {1041–1049},
    numpages = {9},
    location = {Mountain View, California, USA},
    series = {MM '17}
}

@article{Waskom2021Seaborn,
  title={Seaborn: statistical data visualization},
  author={Waskom, Michael L},
  journal={Journal of open source software},
  volume={6},
  number={60},
  pages={3021},
  year={2021}
}

@article{Williams2015Hidden,
    title = {Hidden keys to survival: the type, density, pattern and functional role of emperor penguin body feathers},
    author = {Williams, Cassondra L and Hagelin, Julie C and Kooyman, Gerald L},
    journal = {Proceedings of the Royal Society B: Biological Sciences},
    volume = {282},
    number = {1817},
    pages = {20152033},
    year = {2015},
    publisher = {The Royal Society}
}

@inbook{Winkler2020Lyrebirds,
    author = {Winkler, David W. and Billerman, Shawn M. and Lovette, Irby J.},
    editor = {Billerman, Shawn M. and Keeney, Brooke K. and Rodewald, Paul G. and Schulenberg, Thomas S.},
    title = {Lyrebirds ({Menuridae})},
    booktitle = {Birds of the World},
    version = {1.0},
    publisher = {Cornell Lab of Ornithology},
    address = {Ithaca, NY, USA},
    year = {2020},
    doi = {10.2173/bow.menuri1.01}
}

@inbook{Winkler2020Cotingas,
  author       = {Winkler, D. W. and Billerman, S. M. and Lovette, I. J.},
  title        = {Cotingas ({Cotingidae})},
  booktitle    = {Birds of the World},
  editor       = {Billerman, S. M. and Keeney, B. K. and Rodewald, P. G. and Schulenberg, T. S.},
  year         = {2020},
  version      = {1.0},
  publisher    = {Cornell Lab of Ornithology},
  location     = {Ithaca, NY, USA},
  doi          = {10.2173/bow.coting1.01},
  url          = {https://doi.org/10.2173/bow.coting1.01}
}

@article{Wu2021Genomic,
    title = {Genomic bases underlying the adaptive radiation of core landbirds},
    author = {Wu, Yonghua and Yan, Yi and Zhao, Yuanqin and Gu, Li and Wang, Songbo and Johnson, David H},
    journal = {BMC Ecology and Evolution},
    volume = {21},
    number = {1},
    pages = {162},
    year = {2021},
    publisher = {Springer},
    doi = {10.1186/s12862-021-01888-5},
}

@inproceedings{Zhang2018Fine,
    author = {Zhang, Yabin and Tang, Hui and Jia, Kui},
    title = {Fine-grained visual categorization using meta-learning optimization with sample selection of auxiliary data},
    booktitle = {European Conference on Computer Vision},
    pages = {241--256},
    year = {2018},
    doi = {10.1007/978-3-030-01237-3_15}
}

\end{document}